\pdfoutput=1
\documentclass[amsmath,prd,aps,twocolumn,superscriptaddress,floatfix,nofootinbib]{revtex4}

\usepackage{amsfonts}
\usepackage{bm}
\usepackage{color}
\usepackage{graphicx}
\definecolor{darkgreen}{cmyk}{0.85,0.2,1.00,0.2}

\newcommand{\be}{\begin{equation}}
\newcommand{\ee}{\end{equation}}
\newcommand{\ba}{\begin{eqnarray}}
\newcommand{\ea}{\end{eqnarray}}

\newcommand\lsim{\mathrel{\rlap{\lower4pt\hbox{\hskip1pt$\sim$}}
        \raise1pt\hbox{$<$}}}
\newcommand\gsim{\mathrel{\rlap{\lower4pt\hbox{\hskip1pt$\sim$}}
        \raise1pt\hbox{$>$}}}

\def\n{{\hat{\bf n}}}
\def\k{{\bf k}}
\def\l{{\bf l}}
\def\L{{\bf L}}

\begin{document}

\title{Detecting patchy reionization in the CMB}

\author{Kendrick M.~Smith}
\affiliation{Perimeter Institute for Theoretical Physics, Waterloo ON N2L 2Y5, Canada}
\author{Simone Ferraro}
\affiliation{Berkeley Center for Cosmological Physics, University of California, Berkeley CA 94720, USA}
\affiliation{Miller Institute for Basic Research in Science, University of California, Berkeley CA 94720, USA}

\date{\today}


\begin{abstract}
Upcoming cosmic microwave background (CMB) experiments will measure 
temperature fluctuations on small angular scales with unprecedented precision.
Small-scale CMB fluctuations are a mixture of late-time effects:
gravitational lensing, Doppler shifting of CMB photons by moving electrons
(the kSZ effect), and residual foregrounds.
We propose a new statistic which separates the kSZ signal from the others,
and also allows the kSZ signal to be decomposed in redshift bins.
The decomposition extends to high redshift, and does not require external datasets
such as galaxy surveys.
In particular, the high-redshift signal from patchy reionization can be cleanly
isolated, enabling future CMB experiments to make
high-significance and qualitatively new measurements of the reionization era.
\end{abstract}


\maketitle

\section{Introduction}

On large angular scales ($l \lsim 2000$),
anisotropy in the cosmic microwave background is mainly sourced
by fluctuations at redshift $z \approx 1100$.
On smaller angular scales ($l \gsim 2000$), this ``primary'' anisotropy
is exponentially suppressed, and CMB
fluctuations are mainly a mixture of several ``secondary'' or
late-time effects.

Among secondary effects with the same blackbody spectrum as the primary CMB,
the largest are gravitational lensing, and the kinematic Sunyaev-Zel'dovich (kSZ) effect.
The kSZ effect refers to Doppler shifting of CMB photons as they 
scatter on radially moving inhomogeneities in free electron density \cite{Ostriker:1986fua,Sunyaev:1980vz,Sunyaev:1972eq}.
The kSZ anisotropy can be roughly decomposed into a ``late-time'' contribution 
from redshifts $z \lsim 3$, when inhomogeneities are large due
to gravitational growth of structure, and a ``reionization'' contribution
from redshift $z \sim 7$, when the ionization fraction is expected to be
inhomogeneous during ``patchy'' reionization \cite{Battaglia:2012im,McQuinn:2005ce,Park:2013mv, Alvarez:2015xzu, Zahn:2011vp}.

\begin{figure}[t]
\centerline{\includegraphics[width=7cm]{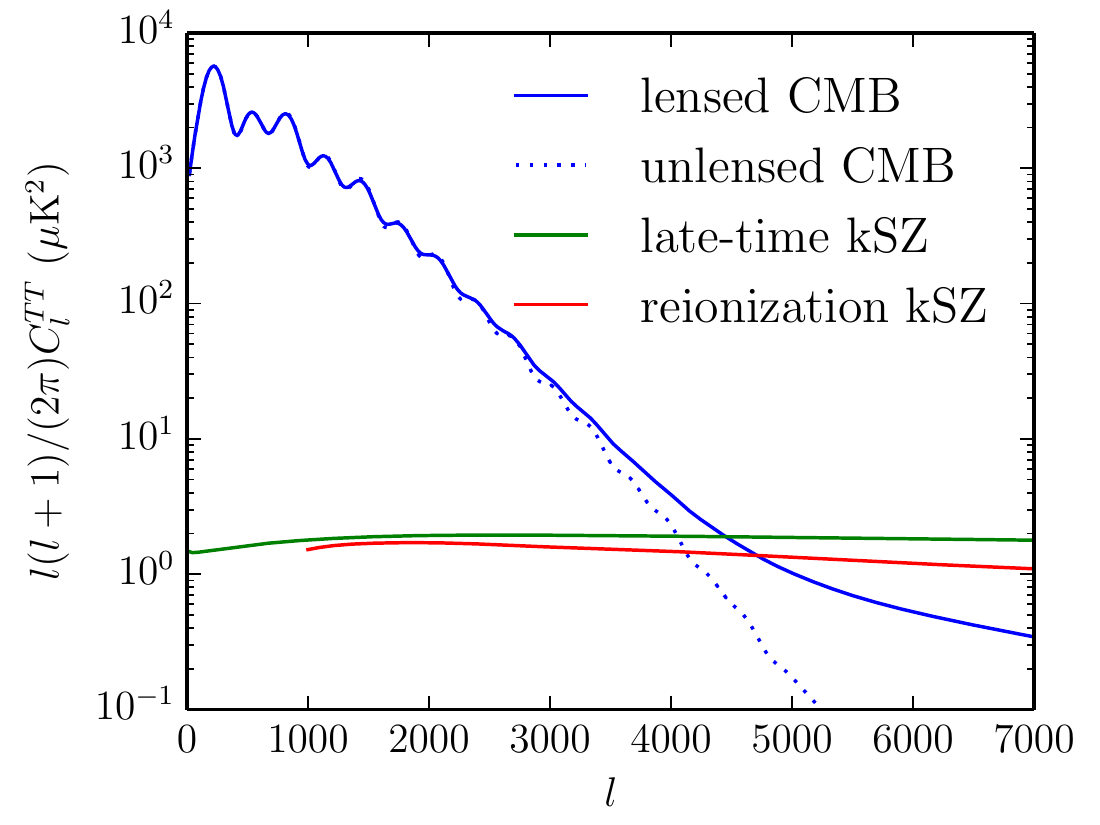}}
\caption{Fiducial model for the CMB temperature power spectrum $C_l^{TT}$
used throughout this paper, split into primary, lensing, late-time kSZ, and
reionization kSZ contributions.}
\label{fig:cltt}
\end{figure}

In Fig.~\ref{fig:cltt} we compare contributions to the temperature power
spectrum $C_l^{TT}$ from weak lensing of the CMB, late-time kSZ, and reionization kSZ,
in a fiducial model to be described shortly.
Individually, these three contributions are very interesting.
Gravitational lensing depends on cosmological parameters such as neutrino mass~\cite{Lewis:2006fu}, 
late-time kSZ probes the distribution of electrons in dark matter halos as well as the large-scale velocity field,
and reionization kSZ may provide the first observational window on 
patchy reionization, which will shed light on the formation of first stars 
and other sources of ionizing photons.
Although the total power spectrum will soon be measured very precisely at high $l$ \cite{Calabrese:2014gwa},
it is unclear how well these signals can be disentangled, since all three components
have large astrophysical modelling uncertainties, and the two kSZ contributions are essentially degenerate at the power spectrum level.

In this paper we will propose a
higher-order statistic which isolates the kSZ signal, and moreover gives
information about its source redshift dependence, allowing the late-time
and reionization kSZ to be separated.  This will complement measurements from future 21cm experiments \cite{Furlanetto:2015apc,Morales:2009gs}. We describe the intuitive
idea here, with a more formal description in the next section.

We first recall that to a good approximation, the kSZ power spectrum may be 
written as an integral \cite{Ma:2001xr}:
\be
C_l^{\rm kSZ} = \int dz \, Q(z) \, \big\langle v_r(z)^2 \big\rangle \, P_e\!\left( \frac{l}{\chi(z)}, z \right)  \label{eq:clksz_integral}
\ee
where $P_e(k,z)$ is the free electron power spectrum,
$\langle v_r^2 \rangle = \langle v^2 \rangle/3$ is the mean squared radial velocity,
and $\chi(z)$ is the comoving distance to redshift $z$.
The radial weight function $Q(z)$ is given by
\be
Q(z) = T_{\rm CMB}^2 \, \frac{H(z)}{\chi(z)^2} \left( \frac{d\bar\tau}{dz} \right)^2 e^{-2\bar\tau(z)}
\ee
where $d\bar\tau/dz$ is the optical depth per unit redshift.

Throughout the paper we will use the following notation frequently.
Let $\bar K$ be the sky-averaged small-scale power spectrum in a fixed high-$l$ band (say $3000 \le l \le 5000$).
For each direction $\n$ on the sky, let $K(\n)$ be the locally measured small-scale power spectrum near sky location $\n$.
The precise definitions of $\bar K$ and $K(\n)$ will be given in the next section.

It is intuitively clear that $K(\n)$
will be better approximated by using the actual realization of $v_r^2(\n,z)$
along the line of sight in direction $\n$ in the integral in Eq.~(\ref{eq:clksz_integral}),
rather than the cosmic average  $\langle v_r^2 \rangle$.
This leads to anisotropy in $K(\n)$ on large angular scales.

To estimate the level of anisotropy,
suppose we divide the line of sight into segments of size 50 Mpc (the coherence length
of the velocity field), and roughly model the radial velocity $v_r$ as
an independent Gaussian random number in every segment.
Since the line of sight is $10^4$ Mpc in length, $K(\n)$ can be roughly
modelled as the sum of squares of $N=200$ independent Gaussians.
This suggests that fluctuations in $K(\n)$ between different lines of sight
are of fractional size $\sqrt{2/N} \approx 0.1$.
Rephrasing, if we measure the CMB in two regions of sky
separated by more than $\sim$1 degree, so that the lines of sight
sample independent realizations of the velocity field,
the kSZ power spectra will differ by $\approx 10\%$.
This is a large non-Gaussian effect which is not present for lensing and
other secondaries, allowing statistical separation of the kSZ signal. 

In fact we can go further by considering $C_L^{KK}$, the angular power spectrum of $K(\n)$.
Suppose we write $C_L^{KK}$ as a sum of contributions from multiple source redshift bins.
In the next section we will show (Fig.~\ref{fig:clkk_model}) that the 
contribution from redshift $z$ has a broad peak at
wavenumber $L_* \sim k_* \chi(z)$, where $k_* \approx 0.01$ $h$ Mpc$^{-1}$.
Thus the shape of $C_L^{KK}$ is source redshift dependent.
In the general case where $C_L^{KK}$ is a sum over redshift bins,
we can ``deconvolve'' the observed $C_L^{KK}$ to infer the contribution
from each bin, thus separating the late-time and reionization kSZ signals.
The main advantage of this method (compared to an analysis based on $C_l^{TT}$) 
is its robustness: we can make statements about reionization which do not depend on
precise modelling of the other contributions.

\section{Modelling the signal}

We use the following fiducial model for the kSZ power spectrum and its source redshift distribution.
We model the late-time kSZ using Eq.~(\ref{eq:clksz_integral}) with $P_e(k,z) = W(k,z)^2 P_{\rm nl}(k,z)$,
where $P_{\rm nl}$ is the nonlinear matter power spectrum from CAMB~\cite{Lewis:1999bs}, and we have defined
\be
W(k,z) = 0.85 \left( 1 + \frac{k D(z)}{0.5 \ h \mbox{ Mpc}^{-1}} \right)^{-1/2}  \label{eq:Wkz}
\ee
where $D(z)$ is the growth function normalized to $D(z) = 1/(1+z)$ at high $z$.
This form of $W(k,z)$ is a simple fitting function which gives approximate agreement with the ``cooling + star formation''
model from~\cite{Shaw:2011sy}, for both $C_l^{\rm kSZ}$ and $dC_l^{\rm kSZ} / dz$.
The prefactor 0.85 assumes that at late times, 15\% of electrons are in the neutral medium or stars~\cite{Fukugita:2004ee}.

We model the reionization kSZ by assigning a Gaussian redshift distribution to the simulated kSZ power spectrum
from Battaglia et al~\cite{Battaglia:2012im}:
\be
\left( \frac{dC_l^{\rm kSZ}}{dz} \right)_{\rm rei} =  \frac{e^{-(z-z_{\rm re})^2/2\sigma_{\rm re}^2}}{(2\pi\sigma_{\rm re}^2)^{1/2}} C_l^{\rm Battaglia}  \label{eq:dcltt_dz_rei}
\ee
where $(z_{\rm re}, \sigma_{\rm re}) = (8.8, 1.0)$.

We now give a formal definition of the quantities $\bar K$ and $K(\n)$ from the introduction.
First fix a filter $W_S(l)$, and define
a high-pass filtered CMB in Fourier space by $T_S(\l) = W_S(\l) T(\l)$.
Unless otherwise specified, we choose $W_S(l) \propto (C_l^{\rm kSZ})^{1/2} / C_l^{\rm tot}$,
where $C_l^{\rm tot}$ is the total CMB power spectrum, including instrumental noise.
We then define $K(\n) = T_S(\n)^2$ by squaring in real space, and define $\bar K$ to be
the all-sky average $\bar K = \langle K(n) \rangle$.  We note that
\be
\bar K = \int \frac{d^2\l}{(2\pi)^2} W_S(l)^2 C_l^{\rm tot}
\ee
so that $\bar K$ can be interpreted as average high-$l$ power (with $W_S^2$-weighting)
and $K(\n)$ can be interpreted as ``locally measured high-$l$ power near $\n$''.

Our main statistic will be $C_L^{KK}$, the ``power spectrum of the power spectrum''.
Note that there are two scales, a small scale $l \gsim 3000$ selected by the filter $W_S$
where the CMB is measured, and a large scale $L \lsim 300$ where clustering in the small-scale
power is measured.
Viewed as a four-point estimator in the CMB, $C_L^{KK}$ sums over
quadruples $T(\l_1) T(\l_2) T(\l_3) T(\l_4)$ which are ``collapsed'', in the sense
that the CMB wavenumbers $|\l_i|$ are large, but the intermediate wavenumber $\L=(\l_1+\l_2)$
is small.
This is similar to CMB lens reconstruction, where the lensing potential
$\phi(\L)$ and its power spectrum $C_L^{\phi\phi}$ are estimated on large scales
using CMB temperature fluctuations on scales $l \gg L$.

Continuing the analogy with lens reconstruction, we define the reconstruction noise
$N_L^{KK}$ to be the value of $C_L^{KK}$ that would be obtained if the small-scale
temperature were a Gaussian field.  A short calculation gives:
\be
N_L^{KK} = 2 \int \frac{d^2\l}{(2\pi)^2} W_S^2(\l) W_S^2(\L-\l) C_l^{\rm tot} C_{\L-\l}^{\rm tot}  \, . \label{eq:nlkk}
\ee
In the regime $L \ll l$ of interest, $N_L^{KK}$ is nearly constant in $L$.

Now we would like to model the effect described in the previous section:
large-scale non-Gaussian contributions to $C_L^{KK}$ due to correlated radial velocities along the line of sight.
We introduce a simple model, the ``$\eta$-model'', as follows.

We write the sky-averaged small-scale power spectrum $\bar K_{\rm kSZ}$ as
an integral $\bar K_{\rm kSZ} = \int dz\, (d\bar K/dz)$, where
\be
\frac{d\bar K}{dz} = \int \frac{d^2\l}{(2\pi)^2} W_S(l)^2 \frac{dC_l^{\rm kSZ}}{dz}
\ee
and our fiducial model for $dC_l^{\rm kSZ}/dz$ was given in
Eqs.~(\ref{eq:clksz_integral}),~(\ref{eq:Wkz}),~(\ref{eq:dcltt_dz_rei}).
The $\eta$-model is the ansatz that, in a fixed realization of the radial 
velocity field $v_r(\n,z)$, the locally measured small-scale power $K(\n)$ 
can be modelled as
\be
K(\n) = \int dz\, \frac{d\bar K}{dz} \eta(\n,z)  \label{eq:K_eta}
\ee
where $\eta(\n,z) = v_r(\n,z)^2 / \langle v_r(z)^2 \rangle$.
In other words, we assume that the locally generated kSZ power along
the line of sight is proportional to $v_r^2$, but neglect additional
non-Gaussian effects.
Fully characterizing kSZ non-Gaussianity on all scales is
outside the scope of this paper.
Here we are simply claiming that at minimum, the non-Gaussian signal 
predicted by Eq.~(\ref{eq:K_eta}) must exist.

In the Limber approximation, the contribution to $C_L^{KK}$ predicted by
the $\eta$-model is
\be
C_L^{KK} = \int dz \, \frac{H(z)}{\chi(z)^2} \left( \frac{d\bar K}{dz} \right)^2 P_\eta^\perp\!\left(\frac{L}{\chi(z)}\right)  \label{eq:clkk_ksz}
\ee
where $P_\eta^\perp(k)$ is the power spectrum of the field $\eta$ evaluated at
a wavenumber $\k$ perpendicular to the line-of-sight direction.
We will compute $P_\eta^\perp$ in linear theory, where it is independent of $z$,
but depends on the direction of $\k$ since $\eta$ is an anisotropic field.
By a short calculation using Wick's theorem, $P_\eta^\perp$ is given by
\be
P_\eta^\perp(k) = \frac{2}{\langle v_r^2 \rangle^2} 
  \int \frac{d^3\k'}{(2\pi)^3} \frac{(k'_r)^2 (k_r-k_r')^2}{k^2 (\k-\k')^2} P_v(k') P_v(\k-\k')
\ee
where $P_v$ is the linear velocity power spectrum
and $k_r=0$ has been assumed.

\begin{figure}[t]
\centerline{\includegraphics[width=7cm]{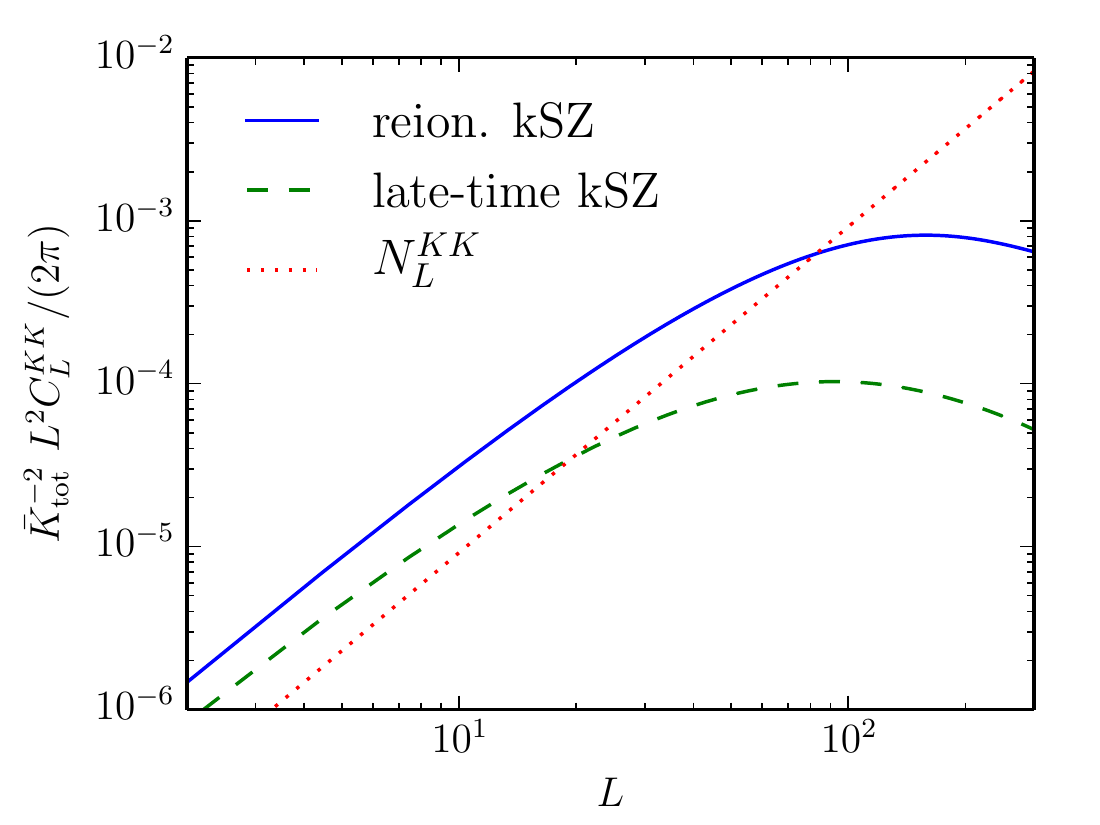}}
\centerline{\includegraphics[width=7cm]{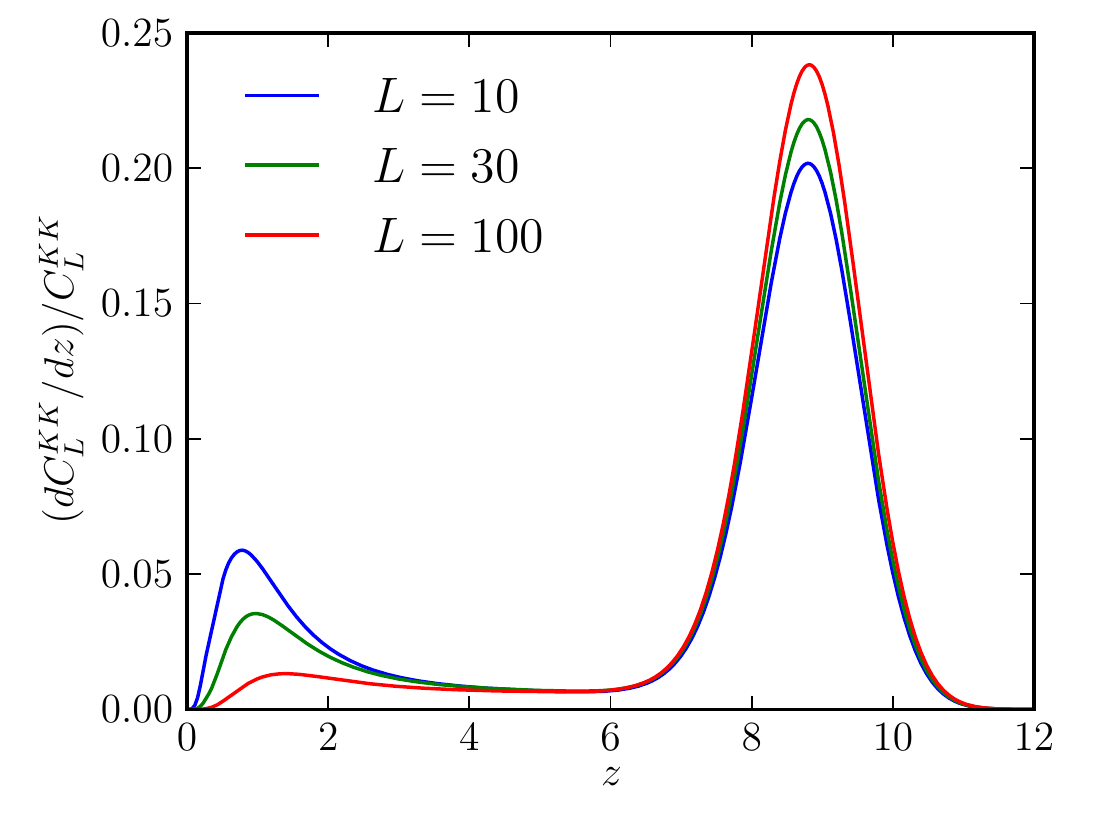}}
\caption{Modelling the ``power spectrum of the kSZ power spectrum'' $C_L^{KK}$,
assuming 2 $\mu$K-arcmin noise and $\theta_{\rm FWHM} = 1'$.
{\em Top panel.} Contributions to $C_L^{KK}$ from reconstruction noise (Eq.~(\ref{eq:nlkk}))
and kSZ (Eq.~(\ref{eq:clkk_ksz})).
{\em Bottom panel.} The kSZ contribution to $C_L^{KK}$ per unit source
redshift.  The distribution is strongly bimodal, justifying a decomposition into low-$z$
and reionization contributions.}
\label{fig:clkk_model}
\end{figure}

In Fig.~\ref{fig:clkk_model} we show the contributions to $C_L^{KK}$ 
from late-time and reionization kSZ, computed using the $\eta$-model.
Note that the reionization kSZ makes a larger contribution to $C_L^{KK}$ than
the late-time kSZ, even though the two are comparable in the CMB
power spectrum $C_l^{TT}$.
This is because the late-time line-of-sight integral is more extended in
comoving distance $\chi$, so it samples more coherence lengths of the velocity
field, making the signal more Gaussian.  

A crucial property of the $\eta$-model is that the contribution to $C_L^{KK}$
from source redshift $z$ is proportional to $P_\eta^\perp(L/\chi(z))$, with no
additional $L$ dependence.
Thus the ``shape'' in $L$ depends only on large-scale linear theory,
but the overall amplitude depends on small-scale physics (via $d\bar K/dz$).

\begin{figure}[t]
\centerline{\includegraphics[width=7cm]{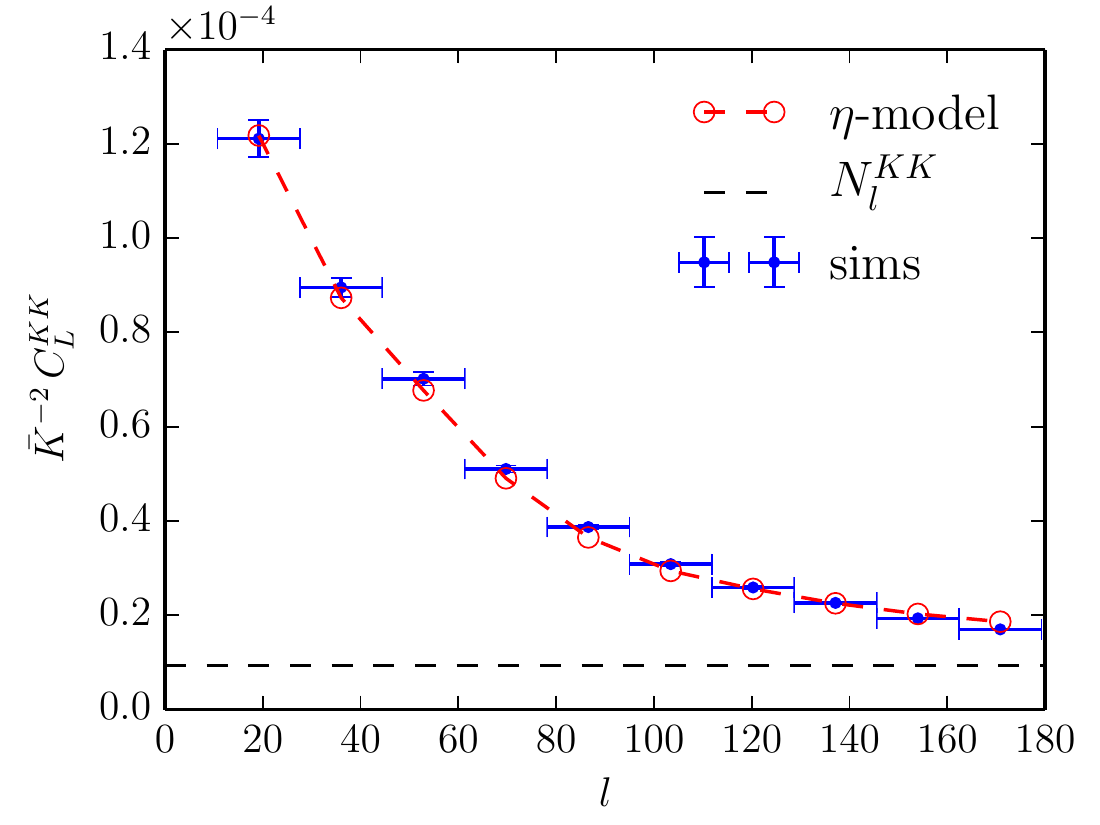}}
\centerline{\includegraphics[width=7cm]{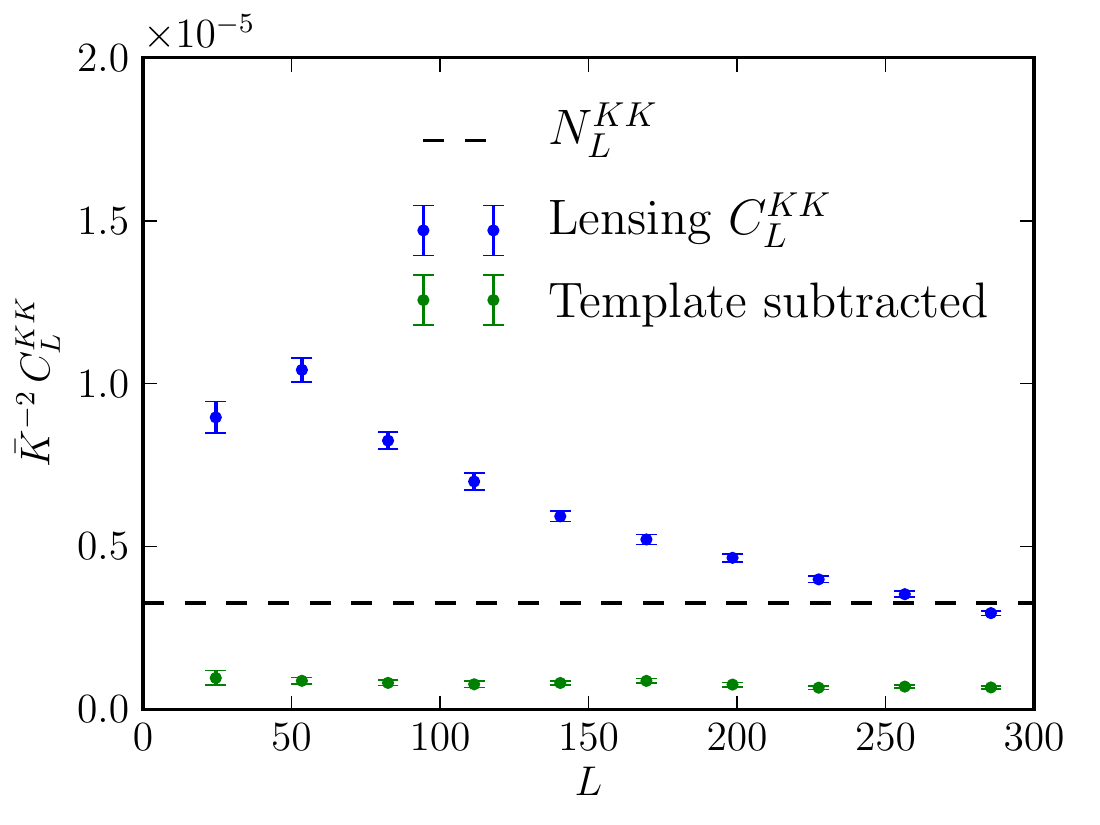}}
\centerline{\includegraphics[width=7cm]{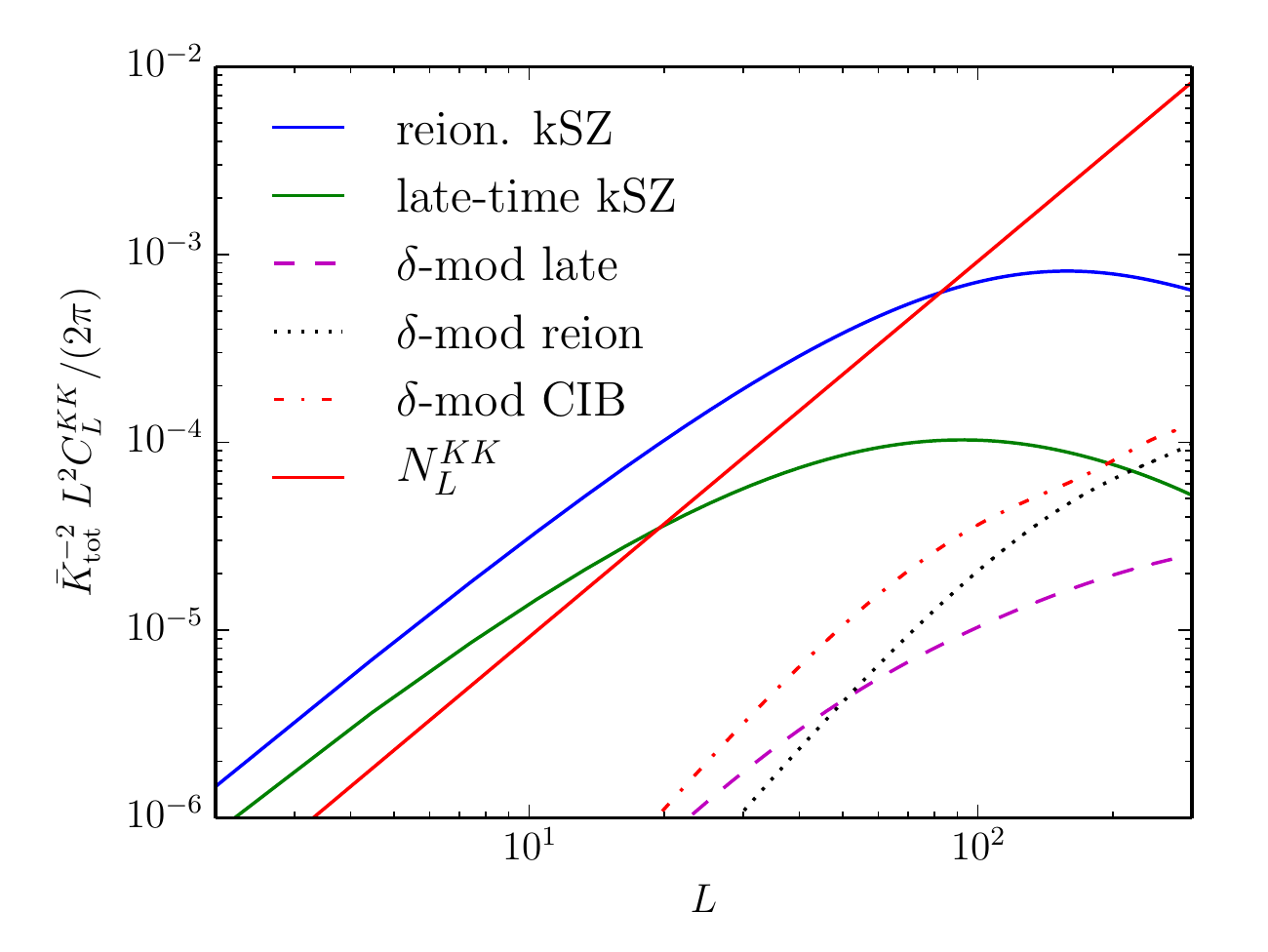}}
\caption{{\em Top panel:} Comparison between our model for $C_L^{KK}$ and 3D large-scale structure simulations
described in the text.  The kSZ maps were filtered to $3000 \le l \le 5000$ with no instrumental noise.
{\em Middle panel:} Lensing contribution to $C_L^{KK}$ from simulations, before
and after template subtraction.
{\em Bottom panel:} Estimated non-Gaussian contribution to $C_L^{KK}$ from large-scale clustering
of small scale modes, computed using a model (``$\delta$-modulation'') described in the text.}
\label{fig:clkk_sims}
\end{figure}

This independence of small-scale physics means that we can test the $\eta$-model
using simplified simulations.  We construct an ensemble of 3D simulations neglecting
baryonic physics and using the 2LPT approximation to the $N$-body equations of motion.
Rather than using a lightcone geometry, we simply project a $z=2$ snapshot onto the 2D
periodic ``sky'' formed by one of the box faces.
The agreement with the $\eta$-model is excellent (Fig.~\ref{fig:clkk_sims} top panel).
We plan to extend this simulation pipeline in future work, but expect that more
accurate simulations will simply change the overall amplitude, and capture small 
effects such as curved-sky corrections and deviations from the Limber approximation.

So far we have considered contributions to $C_L^{KK}$ from the $\eta$-model (our signal),
and from Gaussian mode-counting (the noise $N_L^{KK}$).
In order to claim that the signal is robust, it is important to understand how other 
non-Gaussian effects may contribute to $C_L^{KK}$.

Some non-Gaussian signals do not cluster on large scales.
For example, even after multifrequency analysis, the CMB maps will be contaminated by residual thermal SZ clusters at some level.
This signal can be modelled very accurately as a sum of unclustered Poisson sources with angular profile $F_l$.
On large scales $L$ where the profile $F_L$ is nearly constant, a short calculation
shows that the contribution to $C_L^{KK}$ is nearly constant in $L$.
We will account for this type of contribution by marginalizing an arbitrary constant 
$\delta C_L^{KK}$ in our signal-to-noise forecasts.

Other non-Gaussian signals do cluster on large scales, most importantly gravitational lensing.
In the middle panel of Fig.~\ref{fig:clkk_sims}, we show the bias to $C_L^{KK}$ obtained
from simulated CMB lensing maps, using a pipeline similar to~\cite{BenoitLevy:2012va}.
The lensing bias is comparable to the reconstruction noise and non-constant on large scales.

However, we have found that $C_L^{KK}$ can be ``lens-cleaned'' as follows.
Let us make the ansatz that on large scales, lensing produces terms in $K(\n)$ proportional
to the local lensing convergence $\kappa$ and squared CMB gradient $(\nabla T)^2$:
\be
K(\n) \supset \alpha \kappa(\n) + \beta (\nabla T(\n))^2
\ee
We lens-clean $K$ on large scales by constructing template maps $(\nabla T)^2$ and $\kappa$,
and subtracting best-fit multiples of the templates from $K(\n)$ before computing $C_L^{KK}$.
Our template $(\nabla T)^2$ map is made by low-pass filtering to $\ell < 2000$ and squaring.
We assume that a template map of $\kappa$ is also available in the survey region from CMB
polarization lens reconstruction (a detailed forecast shows that $\kappa$ has high signal-to-noise
for noise levels $\lsim 6$ $\mu$K-arcmin).

Remarkably, this simple template-cleaning procedure removes
nearly all lensing power in simulation (Fig.~\ref{fig:clkk_sims} middle panel).
Furthermore the residual power
is constant in $L$, so that it is removed by our previously mentioned marginalization.  
We therefore expect that CMB lensing bias to $C_L^{KK}$ can be made negligible.

Next we would like to consider clustered secondaries.
One type of non-Gaussian contribution is ``$\delta$-modulated power'':
power along the line of sight whose amplitude is linear in the local overdensity 
$\delta(\n,z)$.
Schematically, we write:
\be
K(\n) \supset \int dz \, \frac{d\bar K}{dz} \Big( 1 + \beta(z) \delta(\n,z) \Big)
\ee
where $\beta(z)$ is a linear bias parameter which relates the small-scale CMB {\em power}
along the line of sight to the local density.
This type of model, in conjunction with the Poisson and Gaussian terms previously considered,
is often used to model large-scale clustering of small-scale 
modes (e.g.~\cite{Takada:2013bfn,Mohammed:2014lja,Baldauf:2015vio}).

We will study $\delta$-modulated power from three sources:
the late-time kSZ, reionization kSZ, and residual CIB.
The model predicts the following contribution to $C_L^{KK}$:
\be
C_L^{KK} = \int dz \, \frac{H(z)}{\chi(z)^2} \beta(z)^2 \left( \frac{\partial\bar K}{\partial z} \right)^2 P_\delta\!\left( \frac{l}{\chi(z)}, z \right)  \label{eq:clkk_clust}
\ee
so in each of the three cases, we will need to know the redshift distribution $d\bar K/dz$
of the small-scale power, and the bias-like parameter $\beta$.
Considering $d\bar K/dz$ first, our fiducial model for the kSZ has already been described, and for the CIB
we conservatively assume residual contamination with total power spectrum equal to the late-time kSZ,
and Gaussian redshift distribution with $(\bar z, \sigma_z) = (2,1)$.

Considering $\beta$ next, assigning precise values would require dedicated 
simulations beyond the scope of this paper, but we will make rough estimates
as follows.
In the limit of high $l$, the kSZ power spectrum is 1-halo dominated 
(or during reionization, 1-bubble dominated).
In this limit, the locally generated kSZ power is simply proportional to
the number density of sources, and therefore the parameter $\beta$ is
equal to the usual linear bias $b$.
For the late-time kSZ and CIB, we will take $\beta \approx 1.5$ as a fiducial value.
For the reionization kSZ, we will take $\beta \approx 6$, a typical bubble bias from 
simulations~\cite{Alvarez:2005sa}.

In Fig.~\ref{fig:clkk_sims}, bottom panel, we show the clustering contributions to $C_L^{KK}$
from $\delta$-modulated late-time kSZ, reionization kSZ, and residual CIB.
In all three cases, the modelling is approximate but should give a rough estimate for the
size of the non-Gaussian clustering effect.  
We find that the clustering terms are small compared to our main signal plus reconstruction noise,
and have fairly different $L$-dependence so that there is little degeneracy.  
This result makes intuitive sense: clustering terms proportional to $P_v$
dominate on large scales, since terms proportional to $P_\delta$ are suppressed by positive
powers of $k$.  It is also consistent with the excellent agreement between the $\eta$-model
and the 3D simulations seen in the top panel of Fig.~\ref{fig:clkk_sims}.

In summary, the $\eta$-model predicts a large kSZ signal in $C_L^{KK}$, that this
prediction agrees with simulations, and that it is robust to a wide range of possible contaminants.

\section{Forecasts and discussion}

In this section, we will present signal-to-noise forecasts.
The first type of forecast we will consider is ``single-bin detection'':
total signal-to-noise of the kSZ $C_L^{KK}$ summed over all source redshifts, 
marginalized over an arbitrary constant $\delta C_L^{KK}$ as previously described.
In our fiducial model, a single-bin detection would get 86\% of its signal-to-noise
from reionization and 14\% from the late-time kSZ.
Therefore a single-bin detection of $C_L^{KK}$ at the expected level would be strong evidence
for patchy reionization in the CMB.

The next observational milestone would be a ``two-bin detection'', in which we fit for
the overall amplitude of the high-$z$ signal ($z \ge 4$), marginalized over an
arbitrary multiple of the low-$z$ contribution ($z \le 4$).
A two-bin detection would measure the amplitude of the patchy reionization signal
without any assumptions on the low-$z$ amplitude.

Finally, we consider a ``three-bin detection'': detection significance of the
$z \ge 5$ contribution, marginalized over independent redshift bins
with $0 \le z \le 2.5$ and $2.5 \le z \le 5$.
A three-bin detection would establish the bimodal redshift dependence of the kSZ sources,
with peaks at late time and during reionization, and little or no power in between, which would also provide a powerful check on systematics.

The precise definitions are as follows:
given $N$ contributions $(\delta C_L^{KK})_1$, $\cdots$, $(\delta C_L^{KK})_N$
to $C_L^{KK}$ such that $C_L^{KK} = \sum_i A_i (\delta C_L^{KK})_i $ (for example $N$ redshift bins), 
we forecast signal-to-noise on the amplitudes $A_i$ by computing the $N$-by-$N$ Fisher matrix
\be
F_{ij} = \frac{f_{\rm sky}}{2} \sum_{L=L_{\rm min}}^{L_{\rm max}} (2L+1) \frac{(\delta C_L^{KK})_i \, (\delta C_L^{KK})_j}{(C_L^{KK})_{\rm tot}^2}  \label{eq:fisher}
\ee
The signal-to-noise of $(\delta C_L^{KK})_i$, marginalizing over signals $j \ne i$,
is given by $S/N = (F^{-1}_{ii})^{-1/2}$.
The significance of our ``$N$-bin detection'', where $N=1,2,3$, is defined to be the
signal-to-noise of the highest redshift bin, taking $(C_L^{KK})_{\rm tot}$ to be the
sum of contributions from the lower redshift bins plus reconstruction noise,
and marginalized over the other redshift bins
plus a contribution of the form $\delta C_L^{KK}=\mbox{constant}$.
The maximum multipole $L_{\rm max}$ in Eq.~(\ref{eq:fisher}) will depend in practice
on the extent to which secondary contributions to $C_L^{KK}$ can be modelled as $L$ increases.
As a fiducial value, we have used $L_{\rm max}=300$ here.

\begin{figure}
\centerline{\includegraphics[width=7cm]{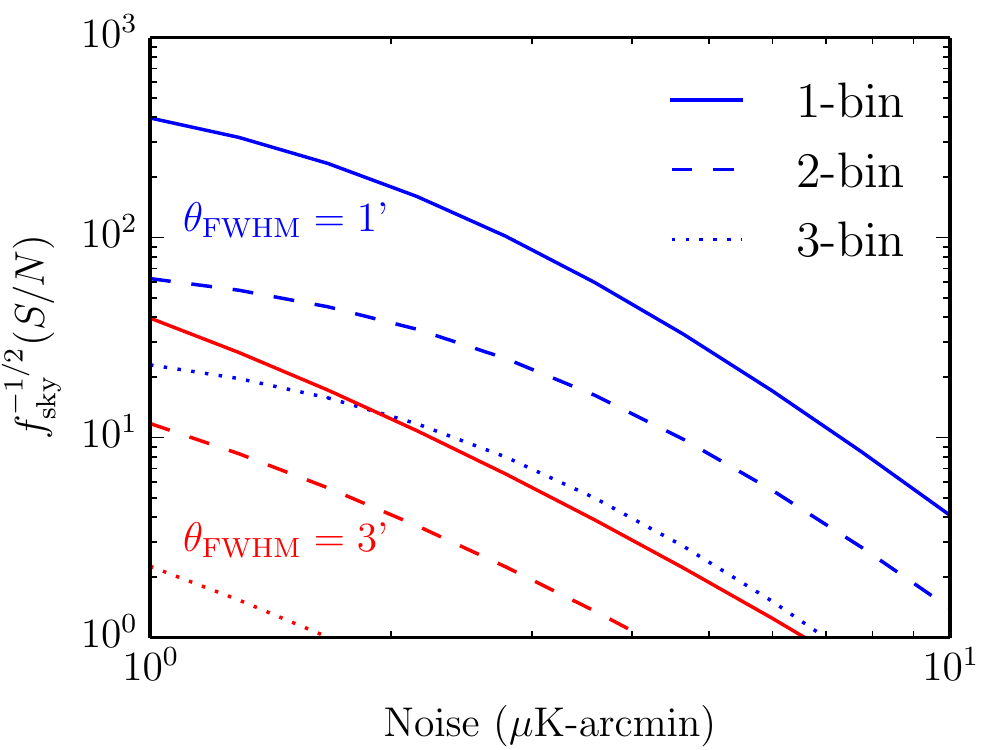}}
\caption{Forecasted $S/N$ for 1-bin, 2-bin, and 3-bin detections as defined in the text,
for varying noise level, and beam size $\theta_{\rm FWHM}=1'$ (blue/upper curves) or
$\theta_{\rm FWHM}=3'$ (red/lower curves).}
\label{fig:forecast}
\end{figure}

In Fig.~\ref{fig:forecast} we show forecasted signal-to-noise as functions of instrumental
noise level and beam size.
These results include improvements from a generalization of the Fisher matrix in Eq.~(\ref{eq:fisher})
in which multiple $K$-fields are defined corresponding to bins in CMB wavenumber $l$.
However, we find that the only case where this significantly improves signal-to-noise
is the three-band detection with $\theta_{\rm FWHM} \lsim 2'$.

It is seen that the signal-to-noise is a steep function of noise level, favoring a deep small-field observing strategy.
However, for surveys smaller than a few hundred square degrees, there is a signal-to-noise
penalty beyond the simple $f_{\rm sky}$ scaling in Fig.~\ref{fig:forecast}, since $C_L^{KK}$
cannot be measured on super-survey scales $L \le L_{\rm min} = (2\pi/\theta_{\rm surv})$.
This signal-to-noise penalty ranges from 5--10\% for a 1000 deg$^2$ survey, and 25--50\% for a 100 deg$^2$ survey,
depending on the noise level and forecast chosen.

Including this penalty, some example surveys which achieve $N$-bin detections are as follows.
A $3\sigma$ one-bin detection can be achieved by a survey with
area $A=500$ deg$^2$, noise $\Delta_T = 4$ $\mu$K-arcmin, and
beam $\theta_{\rm FWHM}=1.4$ arcmin.
Likewise two-bin and three-bin $3\sigma$ detections can be achieved
by surveys with $(A,\Delta_T,\theta_{\rm FWHM}) = (900, 3, 1)$
and $(2400, 2, 1)$ respectively.
An ambitious future survey with 1 $\mu$K-arcmin noise, 1 arcmin beam, and $f_{\rm sky}=0.5$
can achieve 1-bin, 2-bin, and 3-bin detections with significance 279$\sigma$, 44$\sigma$, 
and 16$\sigma$.

In this paper, we have identified a new non-Gaussian signal in the CMB
which is a distinctive observational signature of the kSZ effect.
It should soon be detectable, and an exciting milestone will be a ``clean''
detection of patchy reionization, with minimal assumptions on
modelling of other CMB secondaries. 
Future experiments such as CMB-S4 will have sufficient signal-to-noise to
measure the signal with more granularity and constrain the redshift and
wavenumber dependence of the kSZ sources, opening up a qualitatively new 
observational window on the epoch of reionization.

\vskip 0.2cm

{\em Acknowledgements.}
We thank Joel Meyers and Alex van Engelen for discussions and initial collaboration. We are also grateful to Nick Battaglia, Emmanuel Schaan and David Spergel for discussion.
KMS was supported by an NSERC Discovery Grant and an Ontario Early Researcher Award.
SF was supported by the Miller Institute at the University of California, Berkeley.
Some computations were done on the Scinet cluster at the University of Toronto.
Research at Perimeter Institute is supported by the Government of Canada
through Industry Canada and by the Province of Ontario through the Ministry of Research \& Innovation.

\bibliographystyle{prsty}
\bibliographystyle{h-physrev}
\bibliography{iso_ksz}

\begin{thebibliography}{10}

\bibitem{Ostriker:1986fua}
J.~P. Ostriker and E.~T. Vishniac,
\newblock Astrophys. J. {\bf 306}, L51 (1986).

\bibitem{Sunyaev:1980vz}
R.~A. Sunyaev and {\relax Ya}.~B. Zeldovich,
\newblock Ann. Rev. Astron. Astrophys. {\bf 18}, 537 (1980).

\bibitem{Sunyaev:1972eq}
R.~A. Sunyaev and {\relax Ya}.~B. Zeldovich,
\newblock Comments Astrophys. Space Phys. {\bf 4}, 173 (1972).

\bibitem{Battaglia:2012im}
N.~Battaglia, A.~Natarajan, H.~Trac, R.~Cen, and A.~Loeb,
\newblock Astrophys. J. {\bf 776}, 83 (2013), 1211.2832.

\bibitem{McQuinn:2005ce}
M.~McQuinn, S.~R. Furlanetto, L.~Hernquist, O.~Zahn, and M.~Zaldarriaga,
\newblock Astrophys. J. {\bf 630}, 643 (2005), astro-ph/0504189.

\bibitem{Park:2013mv}
H.~Park {\em et~al.},
\newblock Astrophys. J. {\bf 769}, 93 (2013), 1301.3607.

\bibitem{Alvarez:2015xzu}
M.~A. Alvarez,
\newblock Astrophys. J. {\bf 824}, 118 (2016), 1511.02846.

\bibitem{Zahn:2011vp}
O.~Zahn {\em et~al.},
\newblock Astrophys. J. {\bf 756}, 65 (2012), 1111.6386.

\bibitem{Lewis:2006fu}
A.~Lewis and A.~Challinor,
\newblock Phys. Rept. {\bf 429}, 1 (2006), astro-ph/0601594.

\bibitem{Calabrese:2014gwa}
E.~Calabrese {\em et~al.},
\newblock JCAP {\bf 1408}, 010 (2014), 1406.4794.

\bibitem{Furlanetto:2015apc}
S.~R. Furlanetto,
\newblock (2015), 1511.01131.

\bibitem{Morales:2009gs}
M.~F. Morales and J.~S.~B. Wyithe,
\newblock Ann. Rev. Astron. Astrophys. {\bf 48}, 127 (2010), 0910.3010.

\bibitem{Ma:2001xr}
C.-P. Ma and J.~N. Fry,
\newblock Phys. Rev. Lett. {\bf 88}, 211301 (2002), astro-ph/0106342.

\bibitem{Lewis:1999bs}
A.~Lewis, A.~Challinor, and A.~Lasenby,
\newblock Astrophys. J. {\bf 538}, 473 (2000), astro-ph/9911177.

\bibitem{Shaw:2011sy}
L.~D. Shaw, D.~H. Rudd, and D.~Nagai,
\newblock Astrophys. J. {\bf 756}, 15 (2012), 1109.0553.

\bibitem{Fukugita:2004ee}
M.~Fukugita and P.~J.~E. Peebles,
\newblock Astrophys. J. {\bf 616}, 643 (2004), astro-ph/0406095.

\bibitem{BenoitLevy:2012va}
A.~Benoit-Levy, K.~M. Smith, and W.~Hu,
\newblock Phys. Rev. {\bf D86}, 123008 (2012), 1205.0474.

\bibitem{Takada:2013bfn}
M.~Takada and W.~Hu,
\newblock Phys. Rev. {\bf D87}, 123504 (2013), 1302.6994.

\bibitem{Mohammed:2014lja}
I.~Mohammed and U.~Seljak,
\newblock Mon. Not. Roy. Astron. Soc. {\bf 445}, 3382 (2014), 1407.0060.

\bibitem{Baldauf:2015vio}
T.~Baldauf, U.~Seljak, L.~Senatore, and M.~Zaldarriaga,
\newblock (2015), 1511.01465.

\bibitem{Alvarez:2005sa}
M.~A. Alvarez, E.~Komatsu, O.~Dore, and P.~R. Shapiro,
\newblock Astrophys. J. {\bf 647}, 840 (2006), astro-ph/0512010.

\end{thebibliography}

\end{document}